   \documentstyle[aas2pp4]{article}
   \twocolumn
   \lefthead{GRIV ET AL.}
   \righthead{WEAKLY NONLINEAR THEORY OF THE JEANS INSTABILITY}
   
\begin{document}

\title{Weakly nonlinear theory of the Jeans instability in disk
       galaxies of stars}

\author{Evgeny Griv\altaffilmark{1}, Michael Gedalin, and David Eichler}
\affil{Department of Physics, Ben-Gurion University of the Negev,
       P.O. Box 653, Beer-Sheva 84105, Israel}
\and
\author{Chi Yuan} 
\affil{Academia Sinica Institute of Astronomy and Astrophysics
       (ASIAA), P.O. Box 1-87, Taipei 115, Taiwan}
\altaffiltext{1}{Corresponding author (griv@bgumail.bgu.ac.il).}

\begin{abstract} 

The reaction of collective oscillations excited in the interaction
between aperiodically growing Jeans-type gravity perturbations and
stars of a rapidly rotating disk of flat galaxies is considered.
An equation is derived which describes the change in the main
body of equilibrium distribution of stars in the framework of 
the nonresonant weakly nonlinear theory.
Certain applications of the theory to the problem of relaxation of 
the Milky Way at radii where two-body relaxation is not effective
are explored.  The theory, as applied to the Solar neighborhood,
accounts for observed features, such as the shape for the velocity
ellipsoid of stars and the increase in star random velocities 
with age.

\end{abstract}


\section{Introduction}

Relaxation of stellar distribution in galaxies is not
completely understood yet.  Lynden-Bell (1967) and later Shu (1978)
proposed violent relaxation in spherical-like (that is, nonrotating) 
protogalaxies in not virial equilibrium. The associated relaxation for
an individual to gain or lose energy occurs on the exceedingly short
time much smaller than one typical radial period and well before the
rotating galaxy disk is formed.  The relaxation, however, does not
stop at this stage.  There are numerous observations clearly showing
that there exists ongoing slow relaxation (on the time scale of $10-
20$ rotation periods or even larger) in the rapidly rotating disk of
Milky Way's Galaxy (Wielen 1977; Binney \& Tremaine 1987, p. 470;
Gilmore, King, \& van der Kruit 1990).  This slow relaxation of
the distribution of young stars which were born in the equilibrium
disk of the Galaxy results in a randomization of the velocity
distribution (``Maxwellianization") and a monotonic increase 
of the stellar random velocity dispersion (``heating") from about
15 km$\,$s$^{-1}$ for the youngest stars to about 40 km$\,$s$^{-1}$ 
for the oldest disk stars with increasing stellar ages from $\sim 
10^6$ yr to $\sim 4 \, 10^9$ yr.  Wielen (1977) has found that the
observed increase of the velocity dispersion of disk stars with
increasing age indicates strongly a significant irregular 
gravitational field in the galactic disk.  The irregular field 
causes a rapid diffusion of stellar orbits in velocity 
(and positional) space.  The
nature of this relaxation should be quite different from the violent
relaxation.  Various mechanisms for the slow relaxation have been
proposed.  See, e.g., Grivnev \& Fridman (1990) for a review of the
problem.  In the present
work, we elaborate upon the idea of the collective relaxation:
unstable gravity perturbations in the disk affect the averaged 
velocity distribution function.  The instabilities and subsequent
collective relaxation occur near the equilibrium and the 
perturbations remain relatively small which makes this process
very different from what occurs during violent relaxation.

Apparently, Toomre (1964, p. 1237), Goldreich \& Lynden-Bell (1965),
Barbanis \& Woltjer (1967), Kulsrud (1972), and Jenkins \& Binney 
(1990) have first suggested
instabilities as a cause of enhanced relaxation in disk-shaped
rapidly rotating galaxies.  It was stated that because of its
long-range Newtonian forces a self-gravitating medium (a stellar
``gas," say) would possess collective properties: collective, or
cooperative motions in which all the particles of the system
participate.  These properties would be manifested in the behavior
of small gravity perturbations arising against the equilibrium
background.  Collective processes are analogous to two-body
collisions, except that one particle collides with many which
are collected together by some coherent process such as a wave or
an unstable perturbation.  The collective processes are random, and
usually much stronger than the ordinary two-body collisions
and leads to a random walk of the particles that rapidly takes
the complete system toward thermal equilibrium.  

We present (for the first time as far as we are aware)
a quantitative theory of relatively slow relaxation 
on the Hubble time $\sim 10^{10}$ yr of self-gravitating, rapidly
rotating stellar disks of flat galaxies toward a thermal quasi-steady
state by collective effects.  In the process a star ``collides'' 
with inhomogeneities of a galactic gravitational
field which result from the development of the Jeans instability.
\footnote{The classical Jeans-type instability of small-amplitude
gravity disturbances is one of the most frequent and most important
instabilities in the stellar subsystems of galaxies. 
The Jeans instability is driven by a
strong nonresonant interaction of the gravity fluctuations with 
the bulk of the particle population, and the dynamics of Jeans
perturbations can be characterized as a fluidlike interaction. 
Combined with the familiar Lin--Shu--Kalnajs dispersion relation
this is a venerable suggestion as to why flat galaxies almost always 
exhibit spiral structure (Binney \& Tremaine 1987, p. 336).}  
We find that it successfully accounts for several basic 
observations of the Milky Way.

\section{Equilibrium}

In the rotating frame of a disk galaxy, the collisionless motion
of an ensemble of identical stars in the plane of the system can
be described by the Boltzmann equation for the distribution 
function $f ({\bf r},{\bf v},t)$ without the integral of 
collisions (Lin, Yuan, \& Shu 1969):
   \begin{eqnarray}
\frac{\partial f}{\partial t} + v_r \frac{\partial f}{\partial r}
+ \left( \Omega + \frac{v_\varphi}{r} \right) \frac{\partial f}
{\partial \varphi} + \left(2\Omega v_\varphi + \frac{v_\varphi^2}{r}
+ \Omega^2 r \right. \nonumber \\
- \left. \frac{\partial \Phi}{\partial r} \right)
\frac{\partial f}{\partial v_r} 
- \left( \frac{\kappa^2}{2\Omega} v_r + \frac{v_r v_\varphi}{r}
+ \frac{1}{r} \frac{\partial \Phi}{\partial \varphi} \right)
\frac{\partial f}{\partial v_\varphi} = 0 , \label{eq:kinetic}
   \end{eqnarray}
where $r,\varphi,z$ are the galactocentric cylindrical coordinates,
the total azimuthal velocity of the stars was represented as a sum 
of the random $v_\varphi$ and the basic rotation velocity $V_
{\mathrm{rot}}=r\Omega$, $v_r$ is the velocity in the radial
direction, and the epicyclic frequency $\kappa (r)$ is defined
by $\kappa = 2\Omega [1 + (r/2\Omega)(d\Omega/dr)]^{1/2}$.
The quantity $\Omega(r)$ denotes the angular velocity of galactic
rotation at the distance $r$ from the center, and $\kappa$ varies
from $2\Omega$ for a rigid rotation to $\Omega$ for a Keplerian 
one.  Random velocities are small compared with $r\Omega$.
Collisions are neglected here because the collision frequency 
is much smaller than the cyclic frequency $\Omega$.
In the kinetic equation (1), $\Phi ({\bf r},t)$ is the
total gravitational potential determined self-consistently from the
Poisson equation ${\bf \nabla}^2 \Phi = 4 \pi G \int f d {\bf v} =
4 \pi G n$, where $n$ is the volume density.

The equilibrium state is assumed, for simplicity, to be an
axisymmetric and spatially homogeneous stellar disk.  The
distribution function may also be a function of ${\bf r}$, for 
instance, in the case of an inhomogeneous disk, in which case the
theory is significantly complicated (Alexandrov, Bogdankevich, \&
Rukhadze 1984, p. 425). 
Secondly, in our simplified model, the perturbation is propagating in
the plane of the disk.  This approximation of an infinitesimally thin
disk is a valid approximation if one considers perturbations with a
radial wavelength that is greater than the typical disk thickness.  We
assume here that the stars move in the disk plane so that $v_z=0$. 
This allows us to use the two-dimensional distribution function $f=f
(v_r,v_\varphi,t)\delta(z)$ such that $\int f dv_r dv_\varphi
dz=\sigma$, where $\sigma$ is the surface density.  We expect that the
waves propagating in the disk plane have the greatest influence on the
development of structures in the disk.  The latter suggestion is
strongly supported by numerical simulations (Hohl 1978).

The disk in the equlibrium is described by the following equation:
   \begin{equation}
\left( 2\Omega v_\varphi + \frac{v_\varphi^2}{r} \right)
\frac{\partial f_{\mathrm{e}}}{\partial v_r}
- \left( \frac{\kappa^2}{2\Omega} v_r + \frac{v_r v_\varphi}{r}
\right) \frac{\partial f_{\mathrm{e}}}
{\partial v_\varphi} = 0 , \label{eq:equil}
  \end{equation}
where $\partial f_{\mathrm{e}}/\partial t=0$ and the angular
velocity of rotation $\Omega (r)$ is such that the necessary
centrifugal acceleration is exactly provided by the central
gravitational force $r\Omega^2=\partial \Phi_{\mathrm{e}}/\partial 
r$.  Equation (2) does not determine the equilibrium distribution
$f_{\mathrm{e}}$ uniquely.
For the present analysis we choose $f_{\mathrm{e}}$ in the
form of the anisotropic Maxwellian (Schwartzschild) distribution
  \begin{equation}
f_{\mathrm{e}} = \frac{\sigma_{\mathrm{e}}}{2\pi c_r c_\varphi} 
\exp \left( - \frac{v_r^2}{2c_r^2} 
- \frac{v_\varphi^2}{2c_\varphi^2} \right)
= \frac{2\Omega}{\kappa} \frac{\sigma_{\mathrm{e}}}{2\pi c_r^2}
\exp \left( - \frac{v_\perp^2}{2c_r^2} \right) . \label{eq:maxwell}
   \end{equation}
The Schwarzschild distribution function is a function of the two 
epicyclic constants of motion ${\mathcal{E}}=v_\perp^2/2$ and 
$r_0^2\Omega
(r_0)$, where $r_0 = r + (2\Omega /\kappa^2) v_\varphi$.  These
constants of motion are related to the unperturbed star orbits:
  \begin{eqnarray}
r = - \frac{v_\perp}{\kappa} \left[ \sin (\phi_0 - \kappa t) 
- \sin \phi_0 \right] ; 
v_r = v_\perp \cos \left( \phi_0 - \kappa t \right) ; \nonumber \\
\varphi = \frac{2\Omega}{\kappa} \frac{v_\perp}{r_0 \kappa}
\left[ \cos (\phi_0 - \kappa t) - \cos \phi_0 \right] ; 
v_\varphi \approx r_0 \frac{d\varphi}{dt} \nonumber \\
+ r_0 \frac{v_\perp}{\kappa}\frac{d\Omega}
{dr} \sin \left( \phi_0 - \kappa t \right) 
\approx \frac{\kappa}{2\Omega} v_\perp
\sin \left( \phi_0 - \kappa t \right) , \label{eq:orbits}
   \end{eqnarray}
where $v_\perp$, $\phi_0$ are constants of integration, $v_\perp
/ \kappa r_0 \sim \rho /r_0 \ll 1$, $\rho$ is the mean epicycle
radius, and we follow Lin et al. (1969), Shu (1970), and Griv \&
Peter (1996) making use of expressions for the unperturbed 
epicyclic trajectories of stars in the equilibrium central field
$\Phi_{\mathrm{e}} (r)$. 
In equations (3) and (4), $r_0$ is the radius of the circular
orbit, which is chosen so that the constant of areas for this
circular orbit $r_0^2 (d\varphi_0 /dt)$ is equal to the angular
momentum imtegral $M_z= r^2(d\varphi /dt)$, and $v_\perp^2 = v_r^2 +
(2\Omega / \kappa)^2 v_\varphi^2$.  Also, $\varphi_0$ is the position
angle on the circular orbit, $(d\varphi_0 /dt)^2 = (1/r_0)(\partial
\Phi_{\mathrm{e}} / \partial r)_0 = \Omega^2$. The quantities $\Omega$,
$\kappa$, and $c_r$ are evaluated at $r_0$.  In equation (3) the fact
is used that as follows from equations (4) in a rotating frame the
radial velocity dispersion $c_r$ and the azimuthal velocity dispersion
$c_\varphi$ are connected through $c_r \approx (2\Omega / \kappa)
c_\varphi$.  In the Solar vicinity, $2 \Omega / \kappa \approx 1.7$. 
The distribution function $f_{\mathrm{e}}$ has been normalized according to
$\int_{-\infty}^\infty \int_{-\infty}^\infty f_{\mathrm{e}} d v_r d
v_\varphi = 2\pi (\kappa / 2 \Omega) \int_0^\infty v_\perp d v_\perp
f_{\mathrm{e}}= \sigma_{\mathrm{e}}$, where $\sigma_{\mathrm{e}}$ is the
equilibrium surface density.  Such a distribution function for the
unperturbed system is particularly important because it provides a fit
to observations (Lin et al. 1969; Shu 1970).  It is this equilibrium 
that is examined for stability.

\section{Collisionless Relaxation}
 
We proceed by applying the standard procedure of the weakly nonlinear
(or quasi-linear) approach (Galeev \& Sagdeev 1983; Alexandrov et al. 
1984, p. 408; Krall \& Trivelpiece 1986, p. 512) and decompose the time
dependent distribution function $f=f_0 ({\bf v},t) + f_1 ({\bf v},t)$ 
and the gravitational potential $\Phi =\Phi_0 (r,t)+\Phi_1({\bf
r},t)$ with $|f_1/f_0|\ll 1$ and $|\Phi_1 /\Phi_0| \ll 1$ for all
${\bf r}$ and $t$.  The functions $f_1$ and $\Phi_1$ are 
oscillating rapidly in space and time, while the functions $f_0$ and
$\Phi_0$ describe the slowly developing ``background" against which
small perturbations develop; $f_0(t=0) \equiv f_{\mathrm{e}}$ and
$\Phi_0(t=0) \equiv \Phi_{\mathrm{e}}$.  The distribution $f_0$
continues to distort as long as the distribution is unstable.
Linearizing equation (1) and separating fast and slow varying
variables one obtains:
  \begin{eqnarray}
&& \frac{df_1}{dt} = \frac{\partial \Phi_1}{\partial r}
\frac{\partial f_0}{\partial v_r}
+ \frac{1}{r} \frac{\partial \Phi_1}{\partial \varphi}
\frac{\partial f_0}{\partial v_\varphi} , \label{eq:fast}\\  
&& \frac{\partial f_0}{\partial t} = 
{\big\langle} \frac{\partial \Phi_1}{\partial r} \frac{\partial
f_1}{\partial v_r} + \frac{1}{r} \frac{\partial \Phi_1}
{\partial \varphi} \frac{\partial f_1}
{\partial v_\varphi} {\big\rangle} , \label{eq:slow}
  \end{eqnarray}
where $d/dt$ means the total derivative along the star 
orbit (4) and $\langle \cdots \rangle$ denotes the time
average over the fast oscillations.  To emphasize it again, we
are concerned with the growth or decay of small perturbations
from an equilibrium state.

In the epicyclic approximation, the partial derivatives in
equations (5) and (6) transform as follows (Lin et al 1969;
Shu 1970; Griv \& Peter 1996):
   \begin{eqnarray}
\frac{\partial}{\partial v_r} = v_r \frac{\partial}{\partial 
{\mathcal{E}}} - \frac{2\Omega}{\kappa} \frac{v_\varphi}{v_\perp^2}
\frac{\partial}{\partial \phi_0}; \frac{\partial}{\partial
v_\varphi} = \left( \frac{2\Omega} {\kappa} \right)^2 v_\varphi
\frac{\partial}{\partial {\mathcal{E}}} \nonumber \\
+ \frac{2\Omega}{\kappa}
\frac{v_r}{v_\perp^2} \frac{\partial}{\partial \phi_0} .
\label{eq:partial}
   \end{eqnarray}

To determine oscillation spectra, let us consider the stability
problem in the lowest WKB approximation: the perturbation scale
is sufficiently small for the disk to be regarded as spatially
homogeneous.  This is accurate
for short wave perturbations only, but qualitatively correctly
even for perturbations with a longer wavelength, of the order 
of the disk radius $R$.  In this local WKB approximation in
equations (5) and (6), assuming the weakly inhomogeneous disk,
the perturbation is selected in the form of a plane wave
(in the rotating frame):
   \begin{equation}
f_1, \Phi_1 = \frac{1}{2} \delta f, \delta \Phi \left(
e^{i k_r r + im \varphi - i \omega_* t} + \mathrm{c. \, c.} 
\right) , \label{eq:wkb}
   \end{equation}
where $\delta f$, $\delta \Phi$ are amplitudes that are
constant in space and time, $m$ is the nonnegative azimuthal 
mode number, $\omega_* =\omega - m\Omega$ is the Doppler-shifted
wavefrequency, and $|k_r| R \gg 1$ (Griv \& Peter 1996). 
The solution in such a form
represents a spiral wave with $m$ arms whose shape in the plane
is determined by the relation $k_r (r - r_0) = - m (\varphi -
\varphi_0)$.  With $\varphi$ increasing in the rotation direction, 
we have $k_r > 0$ for trailing spiral patters, which are the most
frequently observed among spiral galaxies.  A change of the sign of
$k_r$ corresponds to changing the sense of winding of the spirals,
i.e., leading ones.  With $m=0$, we have the density waves in the
form of concentric rings that propagate away from the center
when $k_r >0$ or toward the center when $k_r <0$.

In equation (5) using the transformation of the derivatives
$\partial /\partial v_r$ and $\partial /\partial v_\varphi$ given 
by equations (7), one obtains the solution
   \begin{equation}
f_1= \int_{-\infty}^t d t^\prime {\bf v}_\perp
\frac{\partial \Phi_1}{\partial {\bf r}}
\frac{\partial f_0}{\partial {\mathcal{E}}} , \label{eq:f1}
   \end{equation}
where $f_1 (t^\prime = -\infty) \rightarrow 0$.  
In this equation making use of the 
time dependence of perturbations in the form of equation (8) 
and the unperturbed trajectories of stars given by equations (4) 
in the exponential factor, it is straightforward to show that  
   \begin{equation}
f_1=- \Phi_1 (r_0) \frac{\partial f_0}{\partial
{\mathcal{E}}} \sum_{l=-\infty}^\infty \sum_{n=-\infty}^\infty 
l \kappa \frac{e^{i(n-l)(\phi_0 - \zeta)} J_l (\chi) J_n (\chi)}
{\omega_* - l \kappa} , \label{eq:solution}
   \end{equation}
where $J_l (\chi)$ is the Bessel function of the first kind
of order $l$, $\chi =k_*v_\perp /\kappa \sim k_*\rho$, 
$k_* =k\{ 1+[(2\Omega /\kappa)^2 - 1] \sin^2 \psi \}^{1/2}$ 
is the effective wavenumber, $\psi$ is the pitch angle of
perturbations, $\tan \psi = k_\varphi / k_r = m / r k_r$,
and we used the expansion
   \begin{displaymath}
\exp (\pm i b \sin \phi_0) = \sum_{n=-\infty}^\infty
J_n (b) \exp (\pm in\phi_0)
   \end{displaymath} 
and the usual Bessel function recursion relation 
   \begin{displaymath}
J_{l+1} (\chi) + J_{l-1} (\chi) = (2l/\chi) J_l (\chi) .
   \end{displaymath}
In the equation above the denominator
vanishes when $\omega_* - l \kappa \rightarrow 0$.  This occurs
near corotation ($l=0$) and other resonances ($l
= \pm 1, \pm 2, \cdots$).  The Lindblad resonances occur at
radii where the field $(\partial/\partial {\bf r})\Phi_1$
resonates with the harmonics $l=-1$ (inner 
resonance) and $l=1$ (outer resonance) of the epicyclic
(radial) frequency of equilibrium oscillations of stars 
$\kappa$.  Clearly, the location of these resonances
depends on the rotation curve and the spiral pattern speed;
the higher the $m$ value, the closer in radius the resonances
are located (Lin et al. 1969; Shu 1970).  
In this paper, only the main part
of the galactic disk is studied which lies sufficiently far from
the resonances: below in all equations $\omega_* -l\kappa \ne 0$.

We substitute the solution (10) into equation (6). 
Taking into account that the terms $l \ne n$ vanish for axially
symmetric functions $f_0$, after averaging over $\phi_0$ we obtain 
the equation for the slow part of the distribution function:
   \begin{equation}
\frac{\partial f_0}{\partial t} = i \pi \sum_{\bf k} 
\sum_{l=-\infty}^\infty |\Phi_{1,{\bf k}}|^2 \frac{\partial}
{\partial v_\perp} \frac{k_* \kappa}{v_\perp \chi}
\frac{l^2 J_l^2(\chi)}{\omega_* -l \kappa}
\frac{\partial f_0}{\partial v_\perp} . \label{eq:back}
   \end{equation}

As usual in the weakly nonlinear theory, in order to close 
the system one must engage an equation for $\Phi_{1,{\bf k}}$. 
Averaging over the fast oscillations, we have
   \begin{equation}
(\partial / \partial t) |\Phi_{1,{\bf k}}|^2
= 2 \Im \omega_* |\Phi_{1,{\bf k}}|^2 , \label{eq:growth}
   \end{equation}
\noindent
where suffixes $\bf k$ denote the $\bf k$th Fourier component.

Equations (11) and (12) form the closed 
system of weakly nonlinear equations for Jeans oscillations of the 
rotating homogeneous disk of stars, and describe a diffusion in
velocity space.  The spectrum of oscillations and their growth 
rate are (Griv, Rosenstein, Gedalin, \& Eichler 1999a; Griv, 
Gedalin, Eichler, \& Yuan 2000a) 
   \begin{equation}
\frac{k^2 c_r^2}{2 \pi G \sigma_0 |k|} = - \sum_{l=
-\infty}^\infty l\kappa \frac{e^{-x} I_l (x)}{\omega_*-l\kappa} ,
\label{eq:dispersion}
   \end{equation}
\noindent
and $\Im \omega_{*,\mathrm{J}} \approx \sqrt{4\pi G \sigma_0
e^{-x} I(x)/x} \lesssim \Omega$, respectively.  In the Solar  
vicinity, $\Omega \approx 3 \, 10^{-8}$ yr$^{-1}$.  Here, $I_l
(x)$ is a Bessel function of an imaginary argument with its
argument $x \approx k_*^2 \rho^2$ and $\rho = c_r /\kappa$
is now the mean epicyclic radius.  A very important feature
of the instability under consideration is the fact that it is
aperiodic (the real part of the wavefrequency vanishes in a
rotating frame we are using).  Usually the quasi-linear theory 
is applied when the growth rate is small compared with the real
part of the wavefrequency as for the case of the resonance
interaction $\omega = {\bf k} \cdot {\bf v}$, where ${\bf v}$
is the velocity of the particle involved in the interaction.
However, the theory can be applied also to aperiodic instabilities
(Shapiro \& Shevchenko 1963; Alexandrov et al. 1984, p. 420;
Krall \& Trivelpiece 1986, p. 531).  A further simplification
results from restricting the frequency range of the waves
examined by taking the low frequency limit ($|\omega_*|$
less than the epicyclic frequency of any disk star).  In
the opposite case of the high perturbation frequencies,
$|\omega_*| > \kappa$, the effect of the disk rotation
is negligible and therefore not relevant to us.  This is
because in this ``rotationless" case the star motion is
approximately rectilinear on the time and length scales
of interest which are the wave growth/damping periods
and wavelength, respectively (Alexandrov et al 1984, p. 113). 
Thus, the terms in series (10)--(13) for which
$|l| \ge 2$ may be neglected, and consideration will be limited
to the transparency region between the turning points in a disk
(between the inner and outer Lindblad resonances).  
In this case, in equations (10)--(13) the 
function $\Lambda (x) = \exp (-x) I_1 (x)$ starts from $\Lambda
(0)=0$, reaches a maximum $\Lambda_{\mathrm{max}}<1$ at $x \approx
0.5$, and then decreases.  Hence, the growth rate has a maximum at
$x<1$ (see Griv, Yuan, \& Gedalin 1999b, Fig. 2 in their paper).

In general, the growth rate of the Jeans instability is high
$|\Im \omega_{*,\mathrm{J}}| \sim \Omega$; perturbations with
wavelength $\lambda_{\mathrm{J}} \approx 2 \pi \rho$ have the
fastest growth rate (Morozov 1981; Griv \& Peter 1996; Griv et
al. 1999a).  In the Solar vicinity of the Galaxy,
$\lambda_{\mathrm{J}} = 2-4$ kpc.

\section{Astronomical Implications}

As an application of the theory we investigate 
the relaxation of low frequency and Jeans-unstable,
$|\omega_*| < \kappa$ and $\omega_*^2 < 0$, respectively, 
oscillations in the homogeneous galactic disk.
Indeed, already in the 1940s it was
observed that in the Solar neighborhood the random velocity
distribution function of stars with an age $t \gtrsim 10^8$ yr
is close to a Schwarzschild distribution --- a set of Gaussian
distributions along each coordinate in velocity space, i.e.,
close to equilibrium along each coordinate (Chandrasekhar 1960;
Ogorodnikov 1965; Binney \& Tremaine 1987, p. 471).  In
addition, older stellar populations have a higher velocity dispersion
than younger ones.\footnote{The age dependence of velocity dispersions
for stellar populations has always been of particular interest, because
the form of this relationship allows us to judge whether was any
relaxation in the galactic disk and even to determine the mechanism
that was responsible for increase in random star velocities.}
On the other hand, a simple calculation of the relaxation time
of the local disk of the Milky Way due to pairwise star--star 
encounters brings the standard value $\sim 10^{14}$ yr (Chandrasekhar
1960), which considerably exceeds the lifetime of the universe.
According to our approach, collisionless relaxation does play a 
determining role in the evolution of stellar populations of the
Galactic disk.

Evidently, the unstable Jeans oscillations
must influence the distribution function of the main,
nonresonant part of stars in such a way as to hinder the
wave excitation, i.e., to increase the velocity dispersion.
This is because the Jeans instability, being essentially a
gravitational one, tends to be stabilized by random motions 
(Toomre 1964; Shu 1970; Bertin 1980; Morozov 1981; Griv \& 
Peter 1996).  Therefore, along with the growth
of the oscillation amplitude, random velocities must increase at the
expense of circular motion, and finally in the disk there can be
established a quasi-stationary distribution so that the Jeans-unstable
perturbations are completely vanishing and only undamped Jeans-stable
waves remain.\footnote{In turn, the Jeans-stable perturbations are
subject to a resonant Landau-type instability (Griv et al. 2000a).}

In the following, we restrict ourselves to the most 
``dangerous," in the sense of the loss of gravitational stability, 
long-wavelength perturbations, $\chi^2$ and $x^2 \ll 1$ (see
the explanation after eqn. [13]).  Then in equations (10)--(13)  
one can use the expansions $J_1^2 (\chi) \approx
\chi^2/4$ and $e^{-x} I_1 (x) \approx (1/2)x - (1/2)x^2 
+ (5/16)x^3$.  Equation (11) takes the simple form
   \begin{equation}
\frac{\partial f_0}{\partial t} = D \frac{\partial^2 f_0} 
{\partial v_\perp^2} , \label{eq:diffusion}
   \end{equation}
where $D=(\pi /16\kappa^2) \sum_{\bf k} k_*^2 \Im \omega_{*,
\mathrm{J}} |\Phi_{1,{\bf k}}|^2$, $\Im \omega_{*,\mathrm{J}}
> 0$, and both $\Im \omega_{*,\mathrm{J}}$ and
$\Phi_{1,{\bf k}}$ are functions of $t$.  As is seen, the
velocity diffusion coefficient for nonresonant stars $D$
is independent of $v_\perp$ (to lowest order).  This is
a qualitative result of the nonresonant character of the
star's interaction with collective aggregates.

By introducing the standard definitions $d\tau/dt = D(t)$ 
and $d/dt = (d\tau/dt)(d/d\tau)$, equations (12) and (14)
are rewritten as follows:
   \begin{equation}
\frac{\partial f_0}{\partial \tau} - \frac{\partial^2 f_0} 
{\partial v_\perp^2} = 0 , \quad \frac{\partial D}{\partial 
\tau} = 2 \Im \omega_{*,\mathrm{J}} , \label{eq:d}
   \end{equation}
which has the solution 
   \begin{equation}
f_0 (v_\perp,\tau) = \frac{\mathrm{const}}{\sqrt{\tau}} \exp
\left( - \frac{v_\perp^2}{4 \tau} \right) . \label{eq:relax}
   \end{equation}
(We have taken into account the observations that the
distribution of newly born stars is close to the $\delta$-function
distribution, $f_0 ({\bf v}_\perp,t=0) = \delta ({\bf
v}_\perp)$; Grivnev \& Fridman 1990.)  
As is seen from equation (16), during the
development of the Jeans instability, the
Schwarzschild distribution of random velocities (a Gaussian spread
along $v_r,v_\varphi$ coordinates in velocity space) is established. 
The energy of the oscillation field $\propto \sum_{\bf k}
|\Phi_{1,{\bf k}}|^2$ plays the role of a ``temperature" $T$ in the
nonresonant-particle distribution.  As the perturbation energy
increases, the initially monoenergetic distribution spreads ($f_0$
becomes less peaked), and the effective temperature grows with time 
(a Gaussian spread increases): $T=2\tau \propto \int D(t)dt\propto
\int \sum_{\bf k}k_*^2 \Im \omega_{*,\mathrm{J}} 
|\Phi_{1,{\bf k}}|^2 dt$. 

From the above, this mechanism increases the velocity
dispersion of stars in Milky Way's disk after they are born.
Subsequently, sufficient velocity dispersion prevents the Jeans
instability from occuring.  The ``diffusion" of nonresonant stars
takes place because they gain mechanical (oscillatory) energy as the
instability develops.  The velocity diffusion, however, presumably
tapers off as Jeans stability is approached: the radial velocity
dispersion $c_r$ becomes greater than the critical one 
$c_{r,\mathrm{crit}} \approx (2\Omega / \kappa) c_{\mathrm{T}}$, where
$c_{\mathrm{T}}$ is the well-known Toomre's critical velocity 
dispersion to suppress the instability of axial symmetric gravity
perturbations (Morozov 1981; Griv \& Peter 1996; Griv et al. 1999a).  
Thus, the true time scale for
relaxation in the Milky Way may be much shorter than its standard
value $\sim 10^{14}$ yr for the classical Chandrasekhar--Ogorodnikov
collisional relaxation; it may be of the order
$(\Im \omega_{*,\mathrm{J}})^{-1} \gtrsim \Omega^{-1} \gtrsim 10^9$
yr, i.e., comparable with $10$ periods of the Milky Way rotation in the 
Solar vicinity.  The above relaxation time is in fair agreement with 
both observations (Wielen 1977; Knude, Winther, \& Schnedler-Nielsen
1987; Gilmore et al. 1990; Grivnev \& Fridman 1990; Meusinger,
Stecklum, \& Reimann 1991) and $N$-body simulations of Milky Way's
disk (Hohl 1971; Sellwood \& Carlberg 1984; Griv, Gedalin, Liverts
et al. 2000b).

\acknowledgements

We thank Tzi-Hong Chiueh, Alexei M. Fridman, Peter Goldreich, 
Muzafar N. Maksumov, Shlomi Pistinner, and Raphael Steinitz for 
helpful conversations.  This research was supported by the 
Israel Science Foundation, the Israeli Ministry of Immigrant
Absorption, and the Academia Sinica in Taiwan.  E.G. appreciates
the hospitality of the ASIAA, where the work was begun.

\end{document}